\documentstyle[twoside]{article}
\catcode`\@=11
\long\def\@makefntext#1{
\protect\noindent \hbox to 3.2pt {\hskip-.9pt
$^{{\eightrm\@thefnmark}}$\hfil}#1\hfill}               

\def\@makefnmark{\hbox to 0pt{$^{\@thefnmark}$\hss}}    

\def\ps@myheadings{\let\@mkboth\@gobbletwo
\def\@oddhead{\hbox{}
\rightmark\hfil\eightrm\thepage}
\def\@oddfoot{}\def\@evenhead{\eightrm\thepage\hfil
\leftmark\hbox{}}\def\@evenfoot{}
\def\sectionmark##1{}\def\subsectionmark##1{}}

\oddsidemargin=\evensidemargin
\newcounter{sectionc}\newcounter{subsectionc}\newcounter{subsubsectionc}
\renewcommand{\section}[1] {\vspace{12pt}\addtocounter{sectionc}{1}
\setcounter{subsectionc}{0}\setcounter{subsubsectionc}{0}\noindent
        {\tenbf\thesectionc. #1}\par\vspace{5pt}}
\renewcommand{\subsection}[1] {\vspace{12pt}\addtocounter{subsectionc}{1}
        \setcounter{subsubsectionc}{0}\noindent
        {\bf\thesectionc.\thesubsectionc. {\kern1pt \bfit #1}}\par\vspace{5pt}}
\renewcommand{\subsubsection}[1] {\vspace{12pt}\addtocounter{subsubsectionc}{1}
        \noindent{\tenrm\thesectionc.\thesubsectionc.\thesubsubsectionc.
        {\kern1pt \tenit #1}}\par\vspace{5pt}}
\newcommand{\nonumsection}[1] {\vspace{12pt}\noindent{\tenbf #1}
        \par\vspace{5pt}}

\newcounter{appendixc}
\newcounter{subappendixc}[appendixc]
\newcounter{subsubappendixc}[subappendixc]
\renewcommand{\thesubappendixc}{\Alph{appendixc}.\arabic{subappendixc}}
\renewcommand{\thesubsubappendixc}
        {\Alph{appendixc}.\arabic{subappendixc}.\arabic{subsubappendixc}}

\renewcommand{\appendix}[1] {\vspace{12pt}
        \refstepcounter{appendixc}
        \setcounter{figure}{0}
        \setcounter{table}{0}
        \setcounter{lemma}{0}
        \setcounter{theorem}{0}
        \setcounter{corollary}{0}
        \setcounter{definition}{0}
        \setcounter{equation}{0}
        \renewcommand{\thefigure}{\Alph{appendixc}.\arabic{figure}}
        \renewcommand{\thetable}{\Alph{appendixc}.\arabic{table}}
        \renewcommand{\theappendixc}{\Alph{appendixc}}
        \renewcommand{\thelemma}{\Alph{appendixc}.\arabic{lemma}}
        \renewcommand{\thetheorem}{\Alph{appendixc}.\arabic{theorem}}
        \renewcommand{\thedefinition}{\Alph{appendixc}.\arabic{definition}}
        \renewcommand{\thecorollary}{\Alph{appendixc}.\arabic{corollary}}
        \renewcommand{\theequation}{\Alph{appendixc}.\arabic{equation}}
        \noindent{\tenbf Appendix \theappendixc #1}\par\vspace{5pt}}
\newcommand{\subappendix}[1] {\vspace{12pt}
        \refstepcounter{subappendixc}
        \noindent{\bf Appendix \thesubappendixc. {\kern1pt \bfit #1}}
        \par\vspace{5pt}}
\newcommand{\subsubappendix}[1] {\vspace{12pt}
        \refstepcounter{subsubappendixc}
        \noindent{\rm Appendix \thesubsubappendixc. {\kern1pt \tenit #1}}
        \par\vspace{5pt}}

\newcommand{\textlineskip}{\baselineskip=13pt}
\newcommand{\smalllineskip}{\baselineskip=10pt}

\def\eightcirc{
\begin{picture}(0,0)
\put(4.4,1.8){\circle{6.5}}
\end{picture}}
\def\eightcopyright{\eightcirc\kern2.7pt\hbox{\eightrm c}}

\newcommand{\copyrightheading}[1]
        {\vspace*{-2.5cm}\smalllineskip{\flushleft
        {\footnotesize International Journal of Modern Physics D 
(in press, 1999)}\\
        {\footnotesize $\eightcopyright$\, World Scientific Publishing
         Company}\\
         }}

\def\abstracts#1#2#3{{
        \centering{\begin{minipage}{4.5in}\baselineskip=10pt\footnotesize
        \parindent=0pt #1\par
        \parindent=15pt #2\par
        \parindent=15pt #3
        \end{minipage}}\par}}

\renewenvironment{thebibliography}[1]
        {\frenchspacing
         \ninerm\baselineskip=11pt
         \begin{list}{\arabic{enumi}.}
        {\usecounter{enumi}\setlength{\parsep}{0pt}
         \setlength{\leftmargin 12.7pt}{\rightmargin 0pt} 
         \setlength{\itemsep}{0pt} \settowidth
        {\labelwidth}{#1.}\sloppy}}{\end{list}}

\newcounter{itemlistc}
\newcounter{romanlistc}
\newcounter{alphlistc}
\newcounter{arabiclistc}

\newcommand{\fcaption}[1]{
        \refstepcounter{figure}
        \setbox\@tempboxa = \hbox{\footnotesize Fig.~\thefigure. #1}
        \ifdim \wd\@tempboxa > 5in
           {\begin{center}
        \parbox{5in}{\footnotesize\smalllineskip Fig.~\thefigure. #1}
            \end{center}}
        \else
             {\begin{center}
             {\footnotesize Fig.~\thefigure. #1}
              \end{center}}
        \fi}

\newcommand{\tcaption}[1]{
        \refstepcounter{table}
        \setbox\@tempboxa = \hbox{\footnotesize Table~\thetable. #1}
        \ifdim \wd\@tempboxa > 5in
           {\begin{center}
        \parbox{5in}{\footnotesize\smalllineskip Table~\thetable. #1}
            \end{center}}
        \else
             {\begin{center}
             {\footnotesize Table~\thetable. #1}
              \end{center}}
        \fi}

\def\@citex[#1]#2{\if@filesw\immediate\write\@auxout
        {\string\citation{#2}}\fi
\def\@citea{}\@cite{\@for\@citeb:=#2\do
        {\@citea\def\@citea{,}\@ifundefined
        {b@\@citeb}{{\bf ?}\@warning
        {Citation `\@citeb' on page \thepage \space undefined}}
        {\csname b@\@citeb\endcsname}}}{#1}}

\newif\if@cghi
\def\cite{\@cghitrue\@ifnextchar [{\@tempswatrue
        \@citex}{\@tempswafalse\@citex[]}}
\def\citelow{\@cghifalse\@ifnextchar [{\@tempswatrue
        \@citex}{\@tempswafalse\@citex[]}}
\def\@cite#1#2{{$\null^{#1}$\if@tempswa\typeout
        {IJCGA warning: optional citation argument
        ignored: `#2'} \fi}}

\def\@refcitex[#1]#2{\if@filesw\immediate\write\@auxout
        {\string\citation{#2}}\fi
\def\@citea{}\@refcite{\@for\@citeb:=#2\do
        {\@citea\def\@citea{, }\@ifundefined
        {b@\@citeb}{{\bf ?}\@warning
        {Citation `\@citeb' on page \thepage \space undefined}}
        \hbox{\csname b@\@citeb\endcsname}}}{#1}}

\def\@refcite#1#2{{#1\if@tempswa\typeout
        {IJCGA warning: optional citation argument
        ignored: `#2'} \fi}}

\def\refcite{\@ifnextchar[{\@tempswatrue
        \@refcitex}{\@tempswafalse\@refcitex[]}}

\def\pmb#1{\setbox0=\hbox{#1}
        \kern-.025em\copy0\kern-\wd0
        \kern.05em\copy0\kern-\wd0
        \kern-.025em\raise.0433em\box0}

\def\fnm#1{$^{\mbox{\scriptsize #1}}$}
\def\fnt#1#2{\footnotetext{\kern-.3em
        {$^{\mbox{\scriptsize #1}}$}{#2}}}

\def\fpage#1{\begingroup
\voffset=.3in
\thispagestyle{empty}\begin{table}[b]\centerline{\footnotesize #1}
        \end{table}\endgroup}

\def\runninghead#1#2{\pagestyle{myheadings}
\markboth{{\protect\footnotesize\it{\quad #1}}\hfill}
{\hfill{\protect\footnotesize\it{#2\quad}}}}
\font\tenrm=cmr10
\font\tenit=cmti10
\font\tenbf=cmbx10
\font\bfit=cmbxti10 at 10pt
\font\ninerm=cmr9

\font\eightrm=cmr8

\def\qed{\hbox{${\vcenter{\vbox{                        
   \hrule height 0.4pt\hbox{\vrule width 0.4pt height 6pt
   \kern5pt\vrule width 0.4pt}\hrule height 0.4pt}}}$}}

\begin{document}

\def\be{\begin{equation}}
\def\ee{\end{equation}}


\runninghead{D. V. Ahluwalia}
{On Quantum Nature of Black-Hole Spacetime $\ldots$}
\normalsize\textlineskip
\thispagestyle{empty}
\setcounter{page}{1}

\copyrightheading{}                     

\vspace*{0.88truein}

\fpage{1}

\centerline{\bf ON QUANTUM NATURE OF BLACK-HOLE SPACETIME:}
\centerline{\bf A Possible New Source of Intense Radiation
\fnm{*}\fnt{*}
{This essay received an ``Honorable Mention'' in the 1999
Essay Competition of the Gravity Research Foundation --- Ed.}
}
\vspace*{0.37truein}

\centerline{\footnotesize D. V. AHLUWALIA\fnm{+}\fnt{+}
{E-mail: ahluwalia@phases.reduaz.mx, 
vahluwa@cantera.reduaz.mx}
}
\vspace*{0.015truein}
\baselineskip=10pt
\centerline{\footnotesize\it
Escuela de Fisica, Univ. Aut. de Zacatecas, Apartado Postal C-580}
\vspace*{0.015truein}
\centerline{\footnotesize\it
Zacatecas, ZAC 98068, Mexico}

\vspace*{0.015truein}


\bigskip
\centerline{\footnotesize Communicated by J. Ellis}

\vspace*{0.21truein}

\abstracts{
Atoms and the planets acquire their stability from the  
quantum mechanical incompatibility of the position and momentum 
measurements. This incompatibility is expressed by the fundamental 
commutator $[x,\,p_x]=i\hbar$, or equivalently, via the  Heisenberg's 
uncertainty principle $\Delta x\, \Delta p_x\sim\hbar$. A further 
stability-related phenomenon where the quantum realm plays a dramatic 
role is the collapse of certain stars into white dwarfs and neutron stars. 
Here, an intervention of the Pauli exclusion principle, via the fermionic 
degenerate pressure, stops the gravitational collapse. However, by the 
neutron-star stage the standard quantum realm runs dry. One is
left with the  problematic collapse of a black hole. This essay is devoted 
to a concrete argument on why the black-hole spacetime itself should  
exhibit a quantum nature. The proposed quantum aspect of spacetime
is shown to prevent the general-relativistic dictated  problematic collapse.
The quantum nature of black-hole spacetime is deciphered from a recent 
result on the universal equal-area spacing [$ =\lambda_P^2\, 
4 \ln(3)$]  for black holes. In one interpretation of the emergent picture,
an astrophysical  black hole can fluctuate to 
$\sqrt{\pi/\ln(3)}$ ($ \approx 1.7$) time its classical 
size, and thus allow radiation and matter to escape to the outside observers.
These fluctuations I conjecture provide a new source, perhaps beyond Hawking
radiation, of intense radiation from astrophysical black holes and may be
the primary source of observed radiation from those galactic cores 
what carry black hole(s). The presented interpretation may be used
as a criterion to choose black holes from black hole candidates.}{}{}



\bigskip\noindent
{\bf 1.} For more than two decades theoretical arguments have been 
accumulating in favor of  black holes having a discrete surface area. 
Insightful reviews on the subject include those by 
Ashtekar\cite{Ashtekar} and Bekenstein. \cite{JDB} 
In Ref. [\refcite{JDB}],
in support of the uniform spacing for the discrete area eigenstates,
Bekenstein notes that, {\em ``transition frequencies at large quantum
numbers should equal classical oscillation frequencies'', because a classical
Schwarzschild black hole displays `ringing frequencies' which scale as $M^{-1}$
... This agreement would be destroyed if the area eigenvalues were unevenly
spaced. Indeed, the loop gravity spectrum ... fails this correspondence 
principle test.} Earlier works that obtained uniformly spaced area 
eigenstates for excitations of black holes include string theoretic argument of
Kogan,\cite{Kogan} and the quantum membrane approaches of 
Maggiore\cite{Maggiore} and Lousto\cite{Lousto}, and canonical quantum gravity
approaches of Louko and M\"akel\"a,\cite{LM} among others. 
However, there is no general agreement on the spacing of the area eigenstates,
or on its uniformity.

\medskip
This situation has been changed to an extent. In a recent work Hod has 
argued  that the black-hole uniform area spacing is given by\cite{hod} 
\begin{equation}
\Delta {\cal A}^\prime = \lambda_P^2\, 4  \ln(3)
\qquad\left[ \lambda_P \equiv
\left(\hbar G/c^3\right)^{1/2}, \mbox{the Planck length}
\right],\label{hod}
\end{equation}
Furthermore, Hod has shown that the area spacing $\lambda_P^2\,4 \ln(3)$
is the {\it unique} value consistent with both the area-entropy
thermodynamic relation, with  the  statistical physics
arguments (i.e., with the Boltzmann-Einstein relation), and with the
Bohr's correspondence principle. The {\em universality} of this 
result suggests that there is 
something fundamental that underlies the equal area spacing. This essay is 
an {\it ab initio} attempt towards discerning this hint. I will 
explore the consequences of the conjecture that the uniform area spacing 
is an intrinsic property of the black hole spacetime and holds even before 
the Bohr correspondence limit is reached.

\bigskip\noindent
{\bf 2.}  The simplest black-hole area operator, compatible with the
stated conjecture, resides
in the observation that the  equal energy-level 
spacing of a one-dimensional quantum harmonic oscillator ({\it QHO})
arise  as a consequence of the fundamental 
commutator  $\left[x,\,p_x\right]=i\hbar$ and from  the fact that {\it QHO} 
Hamiltonian operator consists, additively, the operators 
$x^2$ and $p_x^2$. Thus the simplest area operator compatible with the
stated conjecture is
\begin{equation} 
{\cal A} = 2\pi\,\ell _S^2 +  
{2 \left[ \ln(3)\right]^2\over\pi}\,\ell_P^2,\label{area}
\end{equation}
The ``Schwarzschild'' (S) and  the ``Planck'' (P) operators satisfy
the following fundamental commutator 
\begin{equation}
\left[\ell_P,\,\ell_S\right]=i \lambda_P^2. \label{comm}
\end{equation} 
The factor of $2\pi$ in the first term on the right hand side of Eq. 
(\ref{area}) owes its origin to a remark after Eq. (\ref{PS}) below.
\medskip

The stated conjecture thus carries an element of a  unifying 
thread in that it extends the 
quantum-harmonic-oscillator like structure that has been so successful in 
understanding the non-gravitational fields to the black-hole
spacetime itself. It is possible that the two terms on the right hand
side of Eq. (\ref{area})
are connected via a worm hole (whose contribution to the area
operator is presumably negligible and is missing from Eq. 
(\ref{area}), but which provides a traversable throat). The other possibility 
is that quantum fluctuations
of black hole spacetime make the ``compactified dimensions''
physically accessible and  result in the
indicated bifurcation of the area operator. In either case,  
Eq. (\ref{comm}) represents fundamental restriction on the simultaneous
observability of the relevant sectors --- symbolically represented by
``S'' and ``P.'' These speculations on the physical interpretation
have, in part, been inspired by the wormhole papers of 
Kim and Lee,\cite{KL} and
Garattini,\cite{RG} on the one hand, and a large literature that exists
on the non-commutative aspects of spacetime at the Planck scale 
on the other (see Ref. 
[\refcite{JM}] for a brief historical review).  
The latter aspect is an unavoidable consequence of the fact that 
gravitational effects, associated with the quantum measurements of 
spacetime intervals, render the physical spacetime noncommutative.\cite{a1}
\medskip

It is of interest to note that Amelino-Camelia\cite{gac,NV} has recently 
argued that an interferometric gravitational wave detector can be  
used as a quantum-gravity apparatus to probe the fuzzy/foamy picture 
of spacetime. The latter picture of spacetime is a prediction of nearly 
all approaches that hope to combine gravitation and quantum mechanics. 
Thus,  an experimentally accessible quantum-gravity spacetime
emerges in these pictures where one aspect is governed by a 
source-determined length scale, $\lambda_S \simeq 2 G M/c^2$; and the 
other one is governed by the source-independent length 
scale $\lambda_P = \left(\hbar\,G/c^3\right)^{1/2}$. The former aspect 
is associated with the spacetime structure in general relativity, 
while the latter one is a characteristic of some (yet unknown, and
the realm of the stated conjecture) fundamental 
nature of spacetime. The conjecture put forward above incorporates 
both elements in a natural way. As already noted,
the {\it compactified dimensions} that
are central to almost all current theories (strings, membranes, and so on) 
may live as quantum mechanically incompatible dimensions to our 
four dimensional spacetime and are  phenomenologically described by
postulates (\ref{area}) and (\ref{comm}). 

\medskip

Parenthetically,  a note is in place regarding a recent paper by 
Padmanabhan who showed that a modification of the path integral based on the
principle of duality (i.e., invariance of the path integral amplitude
under $ds$, the path segment, going from $ds\,\to\,\lambda_P^2/ds$)
leads to results which are identical to adding a ``zero-point length''
in the spacetime interval.\cite{TP} The equivalence of the duality
invariance, and  that of the transformation of $g_{\mu\nu} dx^\mu dx^\nu
\to g_{\mu\nu} dx^\mu dx^\nu + \lambda_P^2$, is akin to
the postulated premise here.

\medskip
The presented considerations are confined to a Schwarzschild black hole.

\bigskip\noindent
{\bf 3.} It is now natural to introduce the 
quantum operators $d$ and $d^\dagger$ as
\begin{equation}
d={1\over{\lambda_P}}\sqrt{\pi\over{2 \ln(3)}}\left(
\ell_S-i{\ln(3)\over{\pi}}\ell_P\right),\quad
d^\dagger={1\over{\lambda_P}}\sqrt{\pi\over{2 \ln(3)}}\left(
\ell_S+i{\ln(3)\over{\pi}}\ell_P\right).
\end{equation}
{}From the fundamental commutator $\left[\ell_P,\,\ell_S\right]=i\lambda_P^2$
it follows that $\left[d,\,d^\dagger\right]={\mathbf 1}$. Inverting the
above equations for $\ell_S$ and $\ell_P$, and introducing the area
number operator $\eta=d^\dagger d$, the area operator (\ref{area}) 
takes the form
\begin{equation}
{\cal A}=\lambda_P^2 \,4\ln(3)
\left(\eta+{1\over 2}\right).
\end{equation}
The eigenvalues of $\eta$,  
$\eta\vert\eta^\prime\rangle=\eta^\prime\vert\eta^\prime\rangle$,
are zero and positive integers.  As a consequence, the black hole 
area spectrum is quantized
\begin{equation}
{\cal A}^\prime_{\eta^\prime}=
\lambda_P^2\, 4\ln(3)\left(\eta^\prime+{1\over 2}\right)
\qquad \left[\eta^\prime=0,1,2,\ldots\right],\label{zpa}
\end{equation}
and carries with it a fundamental {\it zero-point area}
\begin{equation}
{\cal A}^\prime_0={1\over 2} \,\lambda_P^2\, 4\,\ln(3).
\end{equation}
Since 
\begin{equation}
d\vert\eta^\prime\rangle=\sqrt{\eta^\prime} \,\vert\eta^\prime-1\rangle,\quad
d^\dagger\vert\eta^\prime\rangle
=\sqrt{\eta^\prime+1}\,\vert\eta^\prime+1\rangle,
\end{equation}
the $d$ and $d^\dagger$ obtain  the interpretation of the {\it area
annihilation} and {\it area creation} operators. 

\medskip
The postulated fundamental commutator, and the area operator,
find their initial justification in preventing a black hole from  
collapsing into the  general-relativistic dictated  singularity, while
at the same time reproducing the equal-area spacing.\cite{hod}


\bigskip\noindent
{\bf 4.} It is readily verified that the following relations are valid: 
\begin{equation}
\langle\eta^\prime\vert  
\ell_P^2\vert\eta^\prime\rangle = \lambda_P^2 {\pi\over{\ln(3)}} 
\left(\eta^\prime+{1\over 2}\right), \quad
\langle\eta^\prime\vert  
\ell_S^2\vert\eta^\prime\rangle =
\lambda_P^2 {\ln(3)\over {\pi}} \left(\eta^\prime+{1\over 2}\right).
\label{PS}
\end{equation}
As a result, the area expectation value $\langle\eta^\prime\vert  {\cal A} 
\vert\eta^\prime\rangle = {\cal A}^\prime_{\eta^\prime}$ derives equal 
contribution from the
Schwarzschild and the Planck sectors of the area operator.
Each of these contributions to  ${\cal A}^\prime_{\eta^\prime}$ equals
\begin{equation}
\lambda_P^2 \,2\,\ln(3)\, \left(\eta^\prime+{1\over 2}\right).
\end{equation}
\medskip

Now consider $\eta^\prime \gg 0$, and call such black
holes astrophysical black holes.  In the spirit of  the Bohr's 
correspondence principle, 
I equate $ \lambda_P^2 4\ln(3) \eta^\prime_{astro}$ 
to the classical result, $ 4\pi\lambda_S^2 $. This yields 
\begin{equation}
\eta^\prime_{astro} = {\pi\over{\ln(3)}}\left({\lambda_S\over{\lambda_P}}
\right)^2.
\label{astro}
\end{equation}

\medskip
 With the help of relations (\ref{PS}), 
I arrive at the quantum-gravity uncertainty relation inherent in
the postulated premise
\begin{equation}
\left(\Delta\ell_P\right)_{\eta^\prime}
\left(\Delta\ell_S\right)_{\eta^\prime}\ge\left(\eta^\prime
+{1\over 2}\right)\lambda_P^2
\qquad\left[\eta^\prime=0,1,2,\ldots\right], 
\label{Pl}
\end{equation}
The quantum fluctuations that appears in the 
above-derived  quantum-gravity uncertainty relation, are defined as:
\begin{equation}
\left(\Delta\ell_\zeta\right)_{\eta^\prime}^2 \equiv 
\langle\eta^\prime\vert\ell_\zeta^2\vert \eta^\prime \rangle -
\langle\eta^\prime\vert\ell_\zeta\vert \eta^\prime \rangle^2
\qquad \left[\zeta=S,P\right].
\end{equation}

\medskip
Thus, in the Bohr's correspondence regime, on using Eq. (\ref{astro}), 
it follows that 
\begin{equation}
\left(\Delta\ell_P\right)_{\eta^\prime}
\left(\Delta\ell_S\right)_{\eta^\prime}\ge \left(\pi/ \ln(3)\right)
\lambda_S^2\qquad\left[\eta^\prime=\eta^\prime_{astro}\right].\label{Sch}
\end{equation}
The $\sqrt{\left(\Delta\ell_P\right)_{\eta^\prime} 
\left(\Delta\ell_S\right)_{\eta^\prime}}$ may be interpreted
to provide a rough measure of the quantum size-fluctuations for a 
black hole. With this interpretation, I infer that quantum fluctuations
can allow astrophysical  black holes to fluctuate 
to a size $\sqrt{\pi/\ln(3)}\approx 1.7$ times their classical size. 
This allows radiation and matter to escape to the outside observers.
I know of no reason to discount this as the primary source of intense
radiation from those galactic cores that carry black hole(s).

\medskip
The picture of the quantum-gravity spacetime associated with an
astrophysical black hole that emerges is that of an object with two 
quantum spheres of fluctuations. The one that may be called a 
{\it Schwarzschild sphere,\/} and the other a {\it Planck sphere.\/} 
The  sizes of these two spheres may be characterized by 
[and are obtained on combining Eqs. (\ref{PS}) and 
(\ref{astro})]
\begin{eqnarray}  
\eta^\prime=\eta^\prime_{astro}:&& \nonumber \\
&&\left(\Delta\ell_S\right)_{\eta^\prime}
= \lambda_S, \\
&&\left(\Delta\ell_P\right)_{\eta^\prime}=
\left(\pi/\ln(3)\right)
\,\lambda_S\approx 2.86\,\lambda_S,
\end{eqnarray}
\medskip

As a cautionary remark I note that the vanishing of 
$\langle\eta^\prime\vert\ell_S\vert\eta^\prime\rangle$ 
and 
$\langle\eta^\prime\vert\ell_P\vert\eta^\prime\rangle$ 
does not imply that the associated quantum mechanical length scale
is zero. All it means is that $\ell_S$ and $\ell_P$ should be treated
as (vector) ``momenta'' and ``amplitudes.'' Good measures of the quantum 
length scales of a black hole are  
$\sqrt{\langle\eta^\prime\vert\ell_S^2\vert\eta^\prime\rangle}$ 
and
$\sqrt{\langle\eta^\prime\vert\ell_P^2\vert\eta^\prime\rangle}$.
The latter, because  
$\langle\eta^\prime\vert\ell_S\vert\eta^\prime\rangle=0=
\langle\eta^\prime\vert\ell_P\vert\eta^\prime\rangle$,  
become identical to 
$\left(\Delta\ell_S\right)_{\eta^\prime}$
and
to $\left(\Delta\ell_P\right)_{\eta^\prime}$.

\medskip
That the quantum-gravity space time must have some sort of fuzzy/foamy
structure is not new. What is new, and is the subject of the
present essay, is a precise model of this  fuzzy/foamy structure 
with a predictive and calculational power. Emergence of the zero-point 
area is only one of the physical implications. Interestingly, the
quantum-gravity fluctuations for the low-mass black holes (i.e., for 
$\eta^\prime$ close to zero) are determined by the Planck length 
$ \lambda_P$ as is apparent from an inspection of Eq. (\ref{Pl}). 
On the other hand, as follows from  Eq. (\ref{Sch}), the quantum-gravity 
fluctuations for  astrophysical  black holes are governed by the  
Schwarzschild length, $\lambda_S$.
 
\medskip

For Planck mass black holes if one identifies $\ell_S$ with the
operator $x$, then the uncertainty $\Delta x$, apart
from being restricted by the relation $\Delta x\,\Delta p_x\ge \hbar/2$,
is further constrained to satisfy (according to Eq. \ref{Pl})
\begin{equation}
\Delta x \ge \left(\eta^\prime+{1\over 2}\right) {\lambda_P^2\over
{\left(\Delta\ell_P\right)_{\eta^\prime}}}.
\end{equation}   
Thus, suggesting that the quantum-gravity effects shall require
modification of the fundamental uncertainty relations of the
standard quantum mechanics. Such modifications have already been
suspected and argued for by several authors [\refcite{a1}, 
\refcite{a2}-\refcite{a9}]
and carry important consequences for the theories that
incorporate gravity and quantum mechanics.


\bigskip\noindent
{\bf 6.} I thus conclude that an important physical consequence of the 
conjectured universal equal-area spacing  $\lambda_P^2\, 
4 \ln(3)$  is that the quantum gravity description of the Schwarzschild
black hole spacetimes is characterized by a quantum area operator (Eq.
\ref{area}) and a new fundamental commutator (Eq. \ref{comm}).
The resulting emergence of the zero point area forbids a black hole from its 
collapse into the  general-relativistic dictated singularity.
An interpretation of the Schwarzschild and Planck spheres of 
quantum fluctuations is that they give black holes a size that is 
roughly $\sqrt{\pi/\ln(3)}$ ($ \approx 1.7$) 
times  their classical size. This allows radiation and matter 
to escape from a black hole to an outside observer perhaps 
dramatically beyond the 
Hawking radiation. Apart from astrophysical consequences, the possible 
existence of two mutually incompatible spatial dimensions has important 
cosmological consequences and it has serious impact on  arguments on 
information loss in the context of black holes. Thus,
the interpretation of the universal equal-area spacing  $\lambda_P^2\, 
4 \ln(3)$ as implying a quantum-harmonic-oscillator like structure of
spacetime is the simplest theoretical construct that unifies gravitation
to other non-gravitational interactions in a non trivial manner.
The proposed interpretation carries in it deep seeds for new studies of
the various paradoxes and problems associated with general-relativistic
black holes and cosmology. The conjecture that {\it compactified dimensions} 
which are so central to almost all current theories embedded in 
higher dimensional spacetimes  may live as quantum mechanically 
incompatible dimensions to our 
four dimensional spacetime provides an additional  possibility 
to discern a new physical principle that may constrain the 
compactification procedure dramatically and may even redefine 
compactification itself.

\nonumsection{Acknowledgments}

It is my pleasure thank Giovanni Amelino-Camelia for bringing to my 
attention the above-cited paper of Padmanabhan (and for providing to 
me his comments on an earlier draft of this essay), Jaron Lanier 
for a conversation on the subject, 
Mariana Kirchbach for a critical reading and comments
on the final draft of this essay, and finally
the {\em Gravity Research Foundation} for its continued
encouragement of my work via its annual prizes and
honorable mentions. 

This work was supported by CONACYT (Mexico).

\nonumsection{Note Added}

Reader's attention is directed to Ref. [\refcite{BR}] which appeared
on the {\it LANL archives} after the present essay was accepted 
for publication. Several new references on the gravitationally induced
modification to the uncertainty relations also appeared in print while
this manuscript was under review. The references have been updated 
accordingly.

\vskip 1.25cm

\nonumsection{References}


\vskip 0.75cm


\end{document}